\documentclass[oldversion]{aa}
\usepackage{graphicx}
\usepackage[authoryear]{natbib}
\usepackage{txfonts}
\begin{document}
   \title{Finding a 24\,-day orbital period for the X-ray binary 1A 1118-616}


   \author{R.~Staubert\inst{1}\and
                K.~Pottschmidt\inst{2,3}\and
                V.~Doroshenko\inst{1}\and
                J.~Wilms\inst{4}\and
                S.~Suchy\inst{5}\and
                R.~Rothschild\inst{5}\and
                A.~Santangelo\inst{1}
}

\offprints{staubert@astro.uni-tuebingen.de}

   \institute{Institut f\"ur Astronomie und Astrophysik, Abteilung Astronomie,
     Universit\"at T\"ubingen (IAAT), Sand 1, 72076 T\"ubingen, Germany
  \and
   NASA-Goddard Space Flight Center, Astrophysics Science Division, Code 661, Greenbelt, MD 20771, USA 
   \and
   Center for Space Science and Technology (CRESST), University of Maryland Baltimore County, 1000 Hilltop Circle, 
   Baltimore, MD 21250, USA
   \and
   Dr. Karl Remeis-Sternwarte and Erlangen Center for Astroparticle Physics,
   Universit\"at Erlangen-N\"urnberg, Sternwartstr. 7, 96049 Bamberg, Germany
   \and 
   Center for Astrophysics and Space Sciences (CASS), University of California San Diego, La Jolla, CA 92093-0424, USA
   }


   \date{Received 10.09.2010; accepted 02.11.2010}

\abstract{     
     We report the first determination of the binary period and orbital ephemeris of the Be X-ray binary 
     containing the pulsar 1A~1118-616 (35 years after the discovery of the source). The orbital period is found 
     to be P$_{\rm orb}$ = $24.0 \pm0.4$\,days. The source was observed by \textsl{RXTE} during its last large 
     X-ray outburst in January 2009, which peaked at MJD~54845.4, by taking short 
     observations every few days, covering an elapsed time comparable to the orbital period. Using the 
     phase connection technique, pulse arrival time delays could be measured and an orbital solution 
     determined. The data are consistent with a circular orbit, and the time of 90 degrees longitude was found 
     to be T$_{\pi/2}$ = MJD 54845.37(10), which is coincident with that of the peak X-ray flux. 
  }


   \maketitle

\section{Introduction}
\label{sec:introduction}
The X-ray transient 1A~1118$-$61 was discovered during an outburst in 1974 
by the \textsl{Ariel-5} satellite in the 1.5--30\,keV range \citep{Eyles:1975p3466}.
The modulation with a period of 6.75\,min found in the data used by
\citet{Ives:1975p3476} was initially interpreted as the orbital period of
two compact objects.  \citet{fabian75} then suggested that the observed period 
may be due to a slowly spinning accreting neutron star in a binary system.  
The optical counterpart was identified as the Be-star He 3-640/Wray 793 by
\citet{Chevalier:1975p3481} and classified as an O9.5IV-Ve star with strong
Balmer emission lines and an extended envelope by 
\citet{JanotPacheco:1981p3486}. The distance was estimated to be $5\pm2$\,kpc
\citep{JanotPacheco:1981p3486}. The classification and distance was confirmed
by \citet{Coe:1985p3489} using UV observations of the source.  

\begin{figure}
  \resizebox{\hsize}{!}{\includegraphics[angle=0]{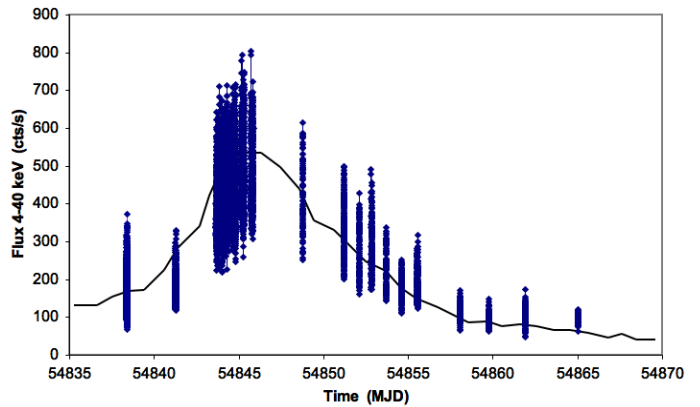}}
  \caption{The light curve of the outburst of 1A~1118-616 as observed
  by \textsl{RXTE}/PCA (4--40\,keV) in January 2009. The time
  resolution is 16\,s. The continuous curve is the daily light curve as 
  seen by \textsl{RXTE}/ASM (scaled to the PCA count rate and 
  smoothed by taking the running mean of five consecutive days).}
  \label{fig:lc}
\end{figure}

A second strong X-ray outburst occurred in January 1992, which was observed by
\textsl{CGRO}/BATSE \citep{Coe:1994p3488}. The measured peak flux was $\sim$
150\,mCrab for the 20-100\,keV energy range, similar to that of the 1974 outburst.
This outburst was followed by enhanced X-ray activity for $\sim$60\,days, with
two flare-like events about $\sim$25\,d and $\sim$49\,d after the peak of the
outburst \citep[see][Fig.~1]{Coe:1994p3488}. Pulsations with $\sim$406.5\,s were 
detected up to 100\,keV and the pulse profile showed a single broad peak, close to 
a sinusoidal modulation with little change with photon energy. A spin-up with a rate of 
0.016\,s/day was observed during the decay of the outburst. Multi-wavelength 
observations revealed a strong correlation between the $H_\alpha$
equivalent width and the X-ray flux. This led \citet{Coe:1994p3488} to
conclude that expansion of the circumstellar disk of the optical companion
is mainly responsible for the increased X-ray activity, including a large
outburst if there is enough matter in the system. This conclusion is supported 
by pulsations from the source also being detected in quiescence 
\citep{quescence}. 

The source remained quiescent until January 4, 2009 when a third outburst was
detected by \textsl{Swift} \citep{Mangano09_GCN8777,swift_atel}. 
Pulsations with a period of $407.68\pm0.02$\,s were reported by
\cite{swift_atel}. The complete outburst was regularly monitored by
\textsl{RXTE}. \textsl{INTEGRAL} observed the source after the main outburst 
\citep{Leyder_etal09}. \textsl{Suzaku} observed 1A~1118$-$61 twice, once during
the peak of the outburst and also $\sim$20\,days later when the flux
returned to its previous level. 
\citet{Dorosh_etal10} analyzed \textsl{RXTE}/PCA data of the most recent outburst
with emphasis on spectral analysis, detecting a cyclotron resonant scattering feature 
(CRSF) at $\sim$55\,keV.

In this letter, we report on an in-depth timing analysis of \textsl{RXTE}/PCA data
over the entire outburst in January 2009, and the discovery of an orbital
period of 24.0\,d. This solution is supported by the timing of the three large
X-ray outbursts seen in history and by the analysis of pulse arrival times of
the January 1992 BATSE observations.

\section{Observations}
We used two data sets. First, and most importantly, we analyzed data obtained 
by  \textsl{RXTE}/PCA during a 27\,-day monitoring of 1A~1118-616, which started 
on January 7, 2009, covering the complete outburst. The  resulting light curve 
(4--40\,keV) is shown in Fig.~\ref{fig:lc}. The source flux peaked on 
January 14 and decayed over $\sim$15\,days. The data structure is defined
by the orbits of the \textsl{RXTE} satellite, with 29 \textsl{RXTE} pointings
of a typical duration of around one hour.
A total exposure of 87\,ks was obtained. The PCA data were all taken in 
\textsl{event  mode} ("Good Xenon"), providing arrival times for individual events.
The data were reduced using HEASOFT version 6.8. 

A second data set consists of pulse profiles generated from archival data of the
observation of the January 1992 outburst by \textsl{CGRO}/BATSE, covering 
about 12\,days.

\begin{figure}
  \resizebox{\hsize}{!}{\includegraphics[angle=90]{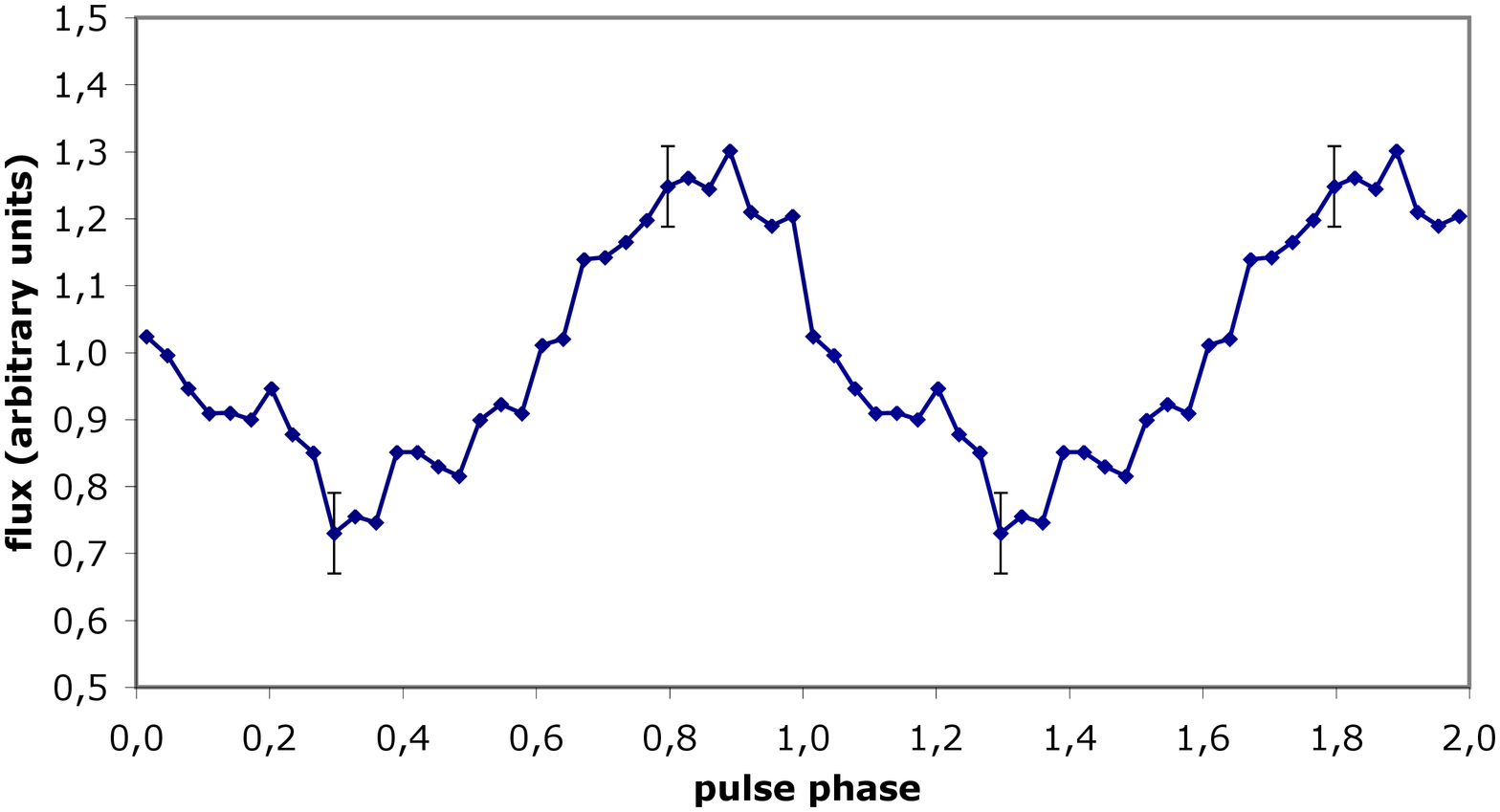}}
  \caption{An example of a pulse profile (with 32 phase bins) of 1A~1118-616 as 
        observed in $\sim$2.5\,ks by \textsl{RXTE}/PCA (normalized to a mean
        flux of 1.0), corresponding to the folding of 6 pulses with a pulse period 
        of $\sim$407.7\,s. The center of the observing time is MJD~54844.375.
        The profile is repeated such that two phases are shown.This profile was 
        used as a template in measuring relative phase shifts of profiles from other 
        integration intervals and the corresponding pulse arrival times.}
  \label{fig:pp}
\end{figure}

\begin{figure}
  \resizebox{\hsize}{!}{\includegraphics[angle=90]{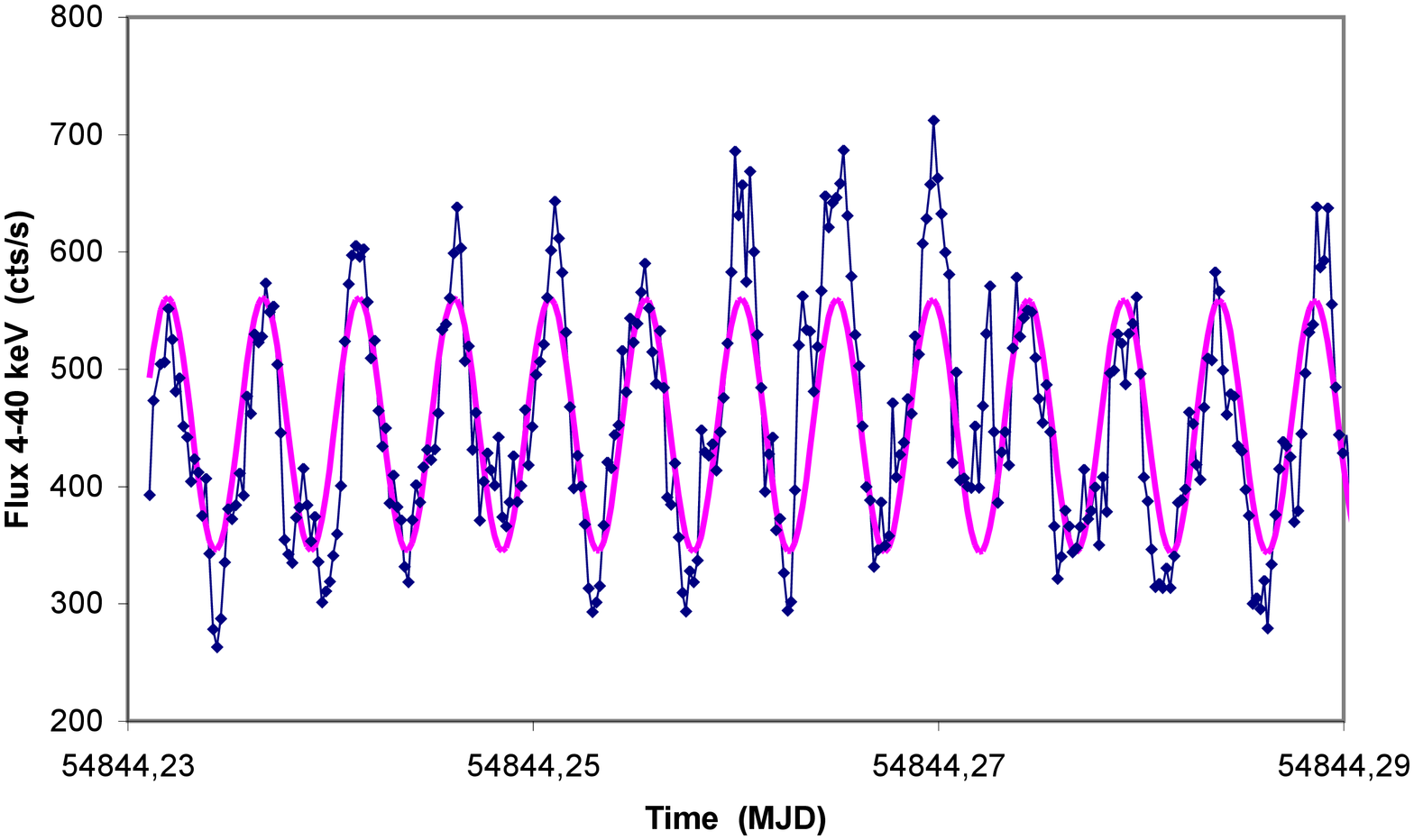}}
  \caption{An example light curve with 16\,s resolution of one
      continuous observation by \textsl{RXTE}/PCA in the
      4--40\,keV interval, together with the best-fit cosine function.}
  \label{fig:sine}
\end{figure}

\begin{table}
\caption{Orbital elements, P$_{\rm pulse}$ and $\dot P_{\rm pulse}$ of 1A~1118-616.
The uncertainties are 1~$\sigma$ (68\%) for two parameters of interest.}
\centering
\begin{tabular}{l l l}
\hline\hline
T$_{\pi/2}$~[MJD (TDB)]              &=& $54845.37 \pm 0.10$ \\
P$_{\rm orb}$ ~[d]                        &=& 24.0  $^{1}$  $\pm$ 0.4 $^{3}$ \\
a~$\sin$~i~[lt-s]                             &=& $54.85 \pm 1.4$ \\
eccentricity $\epsilon$                   &=& 0.0 $^{1}$ \\
$\Omega$~[deg]                                 &=& 360.0 $^{1}$   \\
P$_{\rm pulse}$~[s] at T$_{\rm ref}$$^{2}$   &=& $407.6546 \pm 0.0011$ \\
$\dot P_{\rm pulse}$~[ss$^{-1}$] at T$_{\rm ref}$$^{2}$    &=& $(-1.8\pm 0.2) \times 10^{-7}$ \\
\hline
\end{tabular}
\vspace*{1mm} \\ 
$^{1}$ These parameters were kept fixed (see text), 
$^{2}$ T$_{\rm ref}$ = 54841.259391
$^{3}$ see text regarding this uncertainty
 \label{tab:ephem_ini}
\end{table}

\section{Timing analysis of pulse arrival times \label{timing}}
\subsection{RXTE/PCA - 2009}

Some timing analysis of the \textsl{RXTE} data was performed earlier by \citet{Dorosh_etal10}, 
who determined the pulse period and an initial value of its derivative but failed to 
identify the orbital modulation. Here, a more rigorous analysis is performed, which 
begins in a similar way by selecting 19 integration intervals of the 29 pointings
according to the time structure of the individual satellite pointings (Fig.~\ref{fig:lc}).
These integration intervals ranged from $\sim$1500\,s to $\sim$15000\,s,
corresponding to between four and 38 pulses.
Nineteen pulse profiles (with 32 phase bins) were then constructed by epoch-folding
barycenter-corrected event times (since the Earth is always moving with respect 
to the source, the event arrival times are referenced to the center of the solar system).
Fig.~\ref{fig:pp} shows one of these pulse profiles produced by accumulating 
six pulses (centered at MJD 54844.375).
Using this profile, the \textsl{pulse arrival times} (in MJD) for the other profiles were 
determined by template fitting. The uncertainties of these \textsl{arrival times} are 
of the order of 10\,sec ($\sim$2.5\,\% of the pulse period). 

For the case of no binary modulation and non-zero first and second derivatives of the 
pulse period, the expected \textsl{arrival times} as a function of pulse number $n$ are 
given by (see, e.g., \citealt{Kelley80,Nagase89,staubert2009})
\begin{equation}
t_{\rm n} = t_{\rm 0} + n~P_{\rm s} + \frac{1}{2}~n^{2}~P_{\rm s}\dot P_{\rm s} 
+  \frac{1}{6}~n^{3}~P_{\rm s}^{2}\ddot P_{\rm s} + \dots
\label{cubic}
\end{equation}
With the given accuracy, it was possible to phase-connect the 19 pulse arrival 
times and determine values for the pulse period and its first derivative by
applying a quadratic fit in $n$. The sond derivative
$\ddot P_{\rm s}$ was not constrained and set to zero.
In this fit, significant systematic residuals remained with a sine-like shape and
an amplitude of several tens of seconds, indicative of pulse arrival-time delays
due to orbital motion. Adding the additional term
\begin{equation}
+~a\sin~i \times \cos[2\pi~(t-T_{\pi/2})/P_{\rm orb}]
\label{cos}
\end{equation}
yielded an acceptable fit as expected for a circular binary orbit,
$a\sin i$ being the projected radius of the orbit in light-seconds, and T$_{\pi/2}$ 
(=~T$_{\rm 90}$) in MJD being the time at which the mean orbital longitude of 
the neutron star is $90^\circ$, corresponding to the maximum delay in pulse 
arrival time.
We found that $a\sin i$ $\sim 55$\,light-s, P$_{\rm orb} \sim 24$\,d,
T$_{\rm 90}$ MJD~$\sim 54845$, P$_{\rm pulse}$ $\sim407.7$\,s, and
$\dot P_{\rm s}$ $\sim -1.9\times 10^{-7}$\,ss$^{-1}$.

\begin{figure}
 \resizebox{\hsize}{!}{\includegraphics[angle=90]{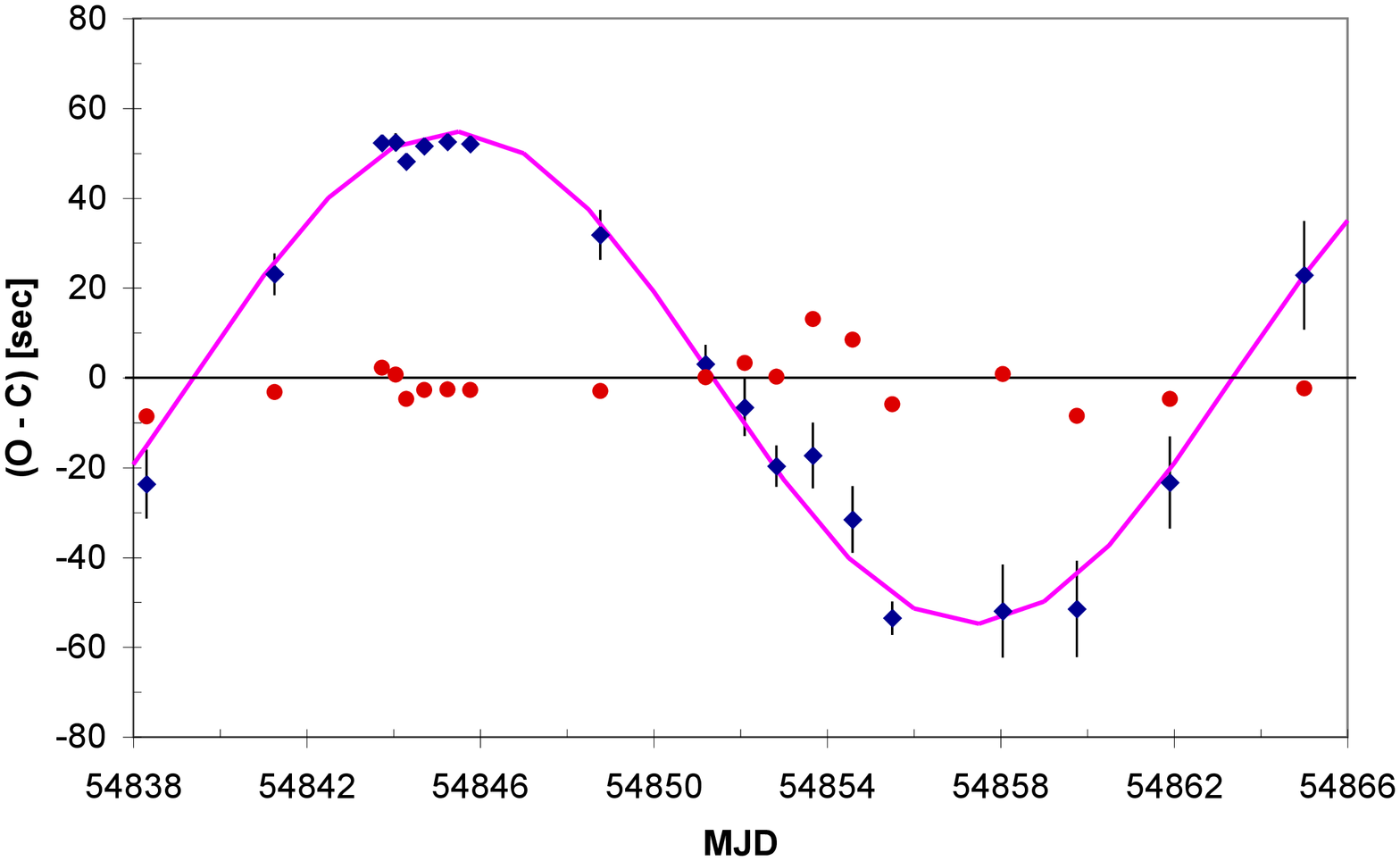}}
  \caption{Delays of the pulse arrival times in 1A~1118-616 for the outburst
     of January 2009 as observed by \textsl{RXTE}/PCA and best-fit sine 
     curve for circular orbital motion with a period of 24.0\,d. 
     The dots around zero are the residuals to the best-fit solution.}
  \label{fig:delay}
\end{figure}

We then realized that by using the original light curve, a sample of which is shown in 
Fig.~\ref{fig:sine}, we were able to determine the \textsl{pulse arrival times} more accurately, 
when fitted piecewise with a cosine function (keeping the pulse period fixed at the previously 
found best-fit values, and leaving the mean flux, the amplitude, and the zero phase as free
parameters). The same 19 integration intervals as before were used.
The phase connection analysis of the \textsl{pulse arrival times} 
determined in this way then leads to the orbital elements and pulse period 
information summarized in Table~\ref{tab:ephem_ini}. The pulse delay times 
and the residuals against the expected delays for the best-fit circular orbit is shown 
in Fig.~\ref{fig:delay}. The $\chi^{2}$ of this best fit was 21.0 for 15\,dof (degrees of
freedom).

In finding this solution, we assumed a vanishing eccentricity and a fixed value
of 24.0\,d for the orbital period. When the orbital period was left as a free parameter,
we found that P$_{\rm orb}$ = $24.0 \pm0.4$\,d. This uncertainty is expected to be large
since our observations cover only 27\,days, little more than one orbital cycle.

Finally, we attempted a proper orbital solution with the eccentricity $\epsilon$   
and the longitude of periastron passage $\Omega$ included as free parameters by 
solving Kepler's equation. The data do allow one to define a pair of parameters with 
reasonably constrained uncertainties of $\epsilon$ = $0.10 \pm0.02$ and 
$\Omega$ = $(310 \pm30)$\,$^{\circ}$. 
However, we consider the evidence of a non-zero eccentricity as marginal. 
By introducing these two additional free parameters, the $\chi^{2}$ was reduced from
21.0 (15\,dof) to 16.4 for (13\,dof). An F-test yielded a rather large probability of 
$\sim$22\,$\%$ for the improvement to have occured by chance. 
Furthermore, no systematic uncertainties of any kind were considered.

We note that the values for the pulse period and its derivative given in Table~1 
differ from those stated by \citet{Dorosh_etal10} because of the missing 
orbital corrections in the earlier analysis.

\begin{figure}
  \resizebox{\hsize}{!}{\includegraphics[angle=90]{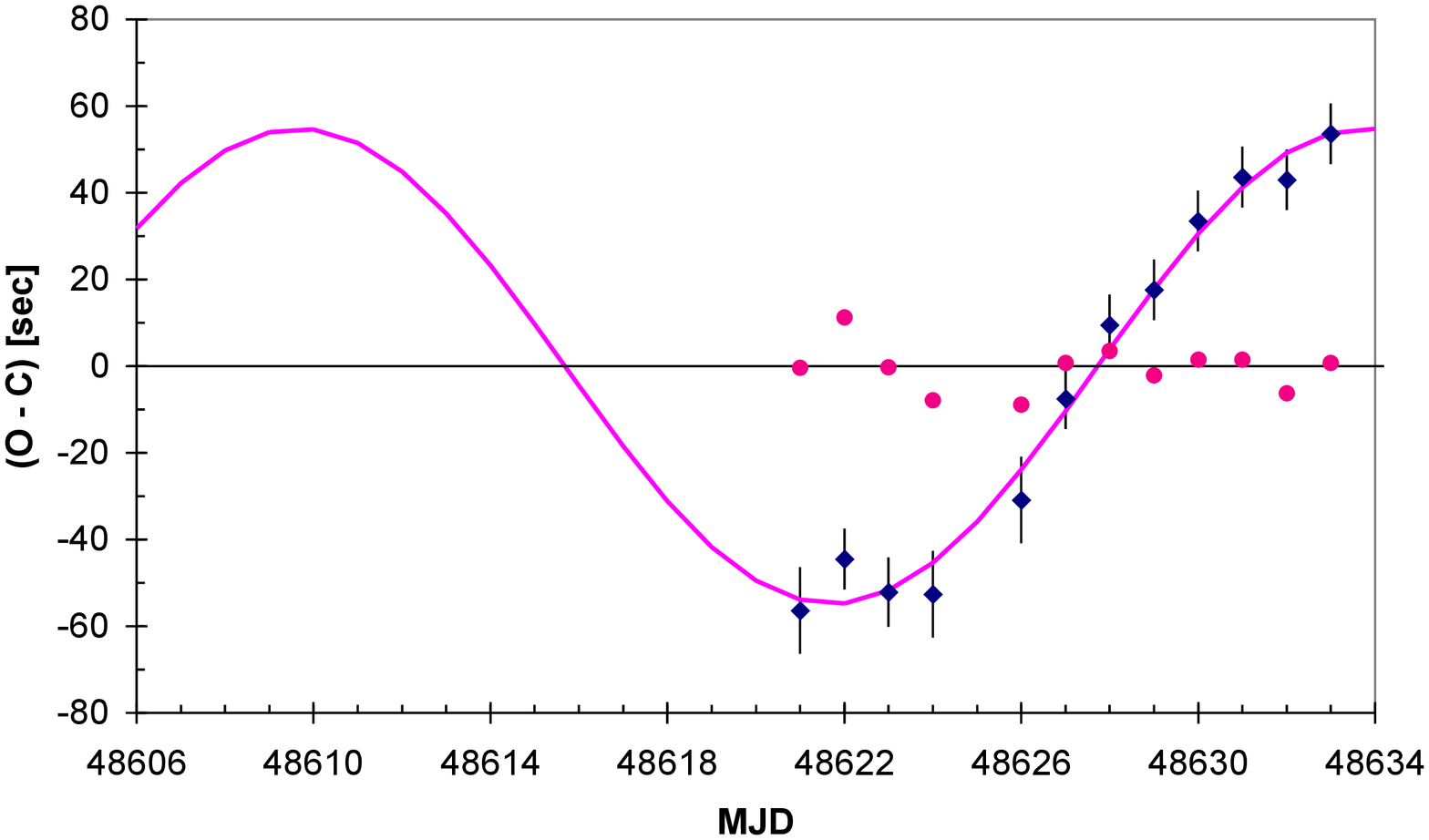}}
  \caption{Delays of the pulse arrival times in 1A~1118-616 for the
   outburst in January 1992 as observed by \textsl{CGRO}/BATSE.
  }
  \label{fig:ar_bat}
\end{figure}

\subsection{CGRO - 1992}

The second data set is from observations of the January 1992 
outburst by \textsl{CGRO}/BATSE. This outburst is described well in 
\citet{Coe:1994p3488}. Pulse profiles were generated using 
phase/energy channel matrices for all eight BATSE detectors which are 
stored at the HEASARC archive.
\footnote{ftp$://$heasarc.gsfc.nasa.gov/compton/data/batse/pulsar/\\
ground folded/A1118-6} 
For each detector, events in the 20--40\,keV energy range (according to the 
energy calibration provided) were selected and sorted into common
pulse profiles with 128 phase bins. This was done for 12 different integration 
intervals, covering about 12\,days of observation (MJD 48621--48633).
Pulse arrival times in MJD were determined by template fitting. 
Fig.~\ref{fig:ar_bat} shows the delays of pulse arrival times as found
from a pulse phase connection analysis. Assuming an orbital period of 
24.0\,d, the best-fit solution for a circular orbit leads to the parameters of 
P$_{\rm pulse}$ = $406.53 \pm0.02$\,s, 
$\dot P_{\rm pulse}$ = $(-3.1\pm 0.9) \times 10^{-7}$ss$^{-1}$, 
and T$_{\pi/2}$ = MJD $48633.5 \pm 2.5$\,d. When the difference between
the T$_{\pi/2}$ value from the \textsl{RXTE} observation (see Table~1)
and the one from the BATSE observation is divided by the orbital period
of 24.0\,d, we find a separation of 258.83 orbital cycles. If, in turn, we
divide the separation by 259 cycles, we determine a cycle length 
of 23.98\,d. The corresponding uncertainty in this value is
0.01\,d. However, we cannot be certain wether the separation
is really 259 cycles, since the previously found uncertainty of 0.4\,d
in the orbital period would permit any cycle number between 255 and
263.

\section{Other support for the 24.0\,d orbital period}

\subsection{Timing of large X-ray outbursts}

Three large X-ray outbursts of 1A~1118-616 have been observed so far. The first one 
occurred in December 1974, leading to the original discovery of the source by \textsl{ARIEL-5} 
\citep{Eyles:1975p3466,Ives:1975p3476}. At this time, a modulation with a period 
of 6.75\,min was also discovered. The second burst was observed by \textsl{CGRO}/BATSE
in January 1992 \citep{Coe:1994p3488}, with enhanced X-ray activity for
$\sim$80\,days. The third, most recent outburst occurred in January 2009 and
was observed by several X-ray satellites: \textsl{Swift} 
\citep{Mangano09_GCN8777,swift_atel}, \textsl{RXTE} \citep{Dorosh_etal10}, 
\textsl{INTEGRAL} \citep{Leyder_etal09}, and \textsl{Suzaku}. 
Also in this case, the source continued to exhibit high activity for about 70\,days after the 
large outburst. From the cited publications, we determined the time of the peak 
fluxes in these outbursts to be MJD~42407.0, MJD~48626.0, and MJD~54845.4, 
respectively, with an estimated uncertainty of $\pm 1$\,day in all cases. 
Taking the above value of 24.0\,d for the orbital period, the outbursts in December 1974
and the one in January 2009 occurred 
259 orbits before and after the one in January 1992.
But, again, because of the uncertainty in P$_{\rm orb}$ the separations could 
range from 255 to 263 orbital cycles, corresponding to a series of period values 
separated by about 0.1\,d. In Section 4.3, however, we present evidence
that 259 may be the correct number. Assuming that this is indeed so, a linear fit 
to the three outburst times leads to a period of $(24.012 \pm0.003)$\,d
(the small uncertainty being due to the long baseline in time). 

\subsection{Timing of medium-size X-ray flares}

\begin{figure}
  \resizebox{\hsize}{!}{\includegraphics[angle=90]{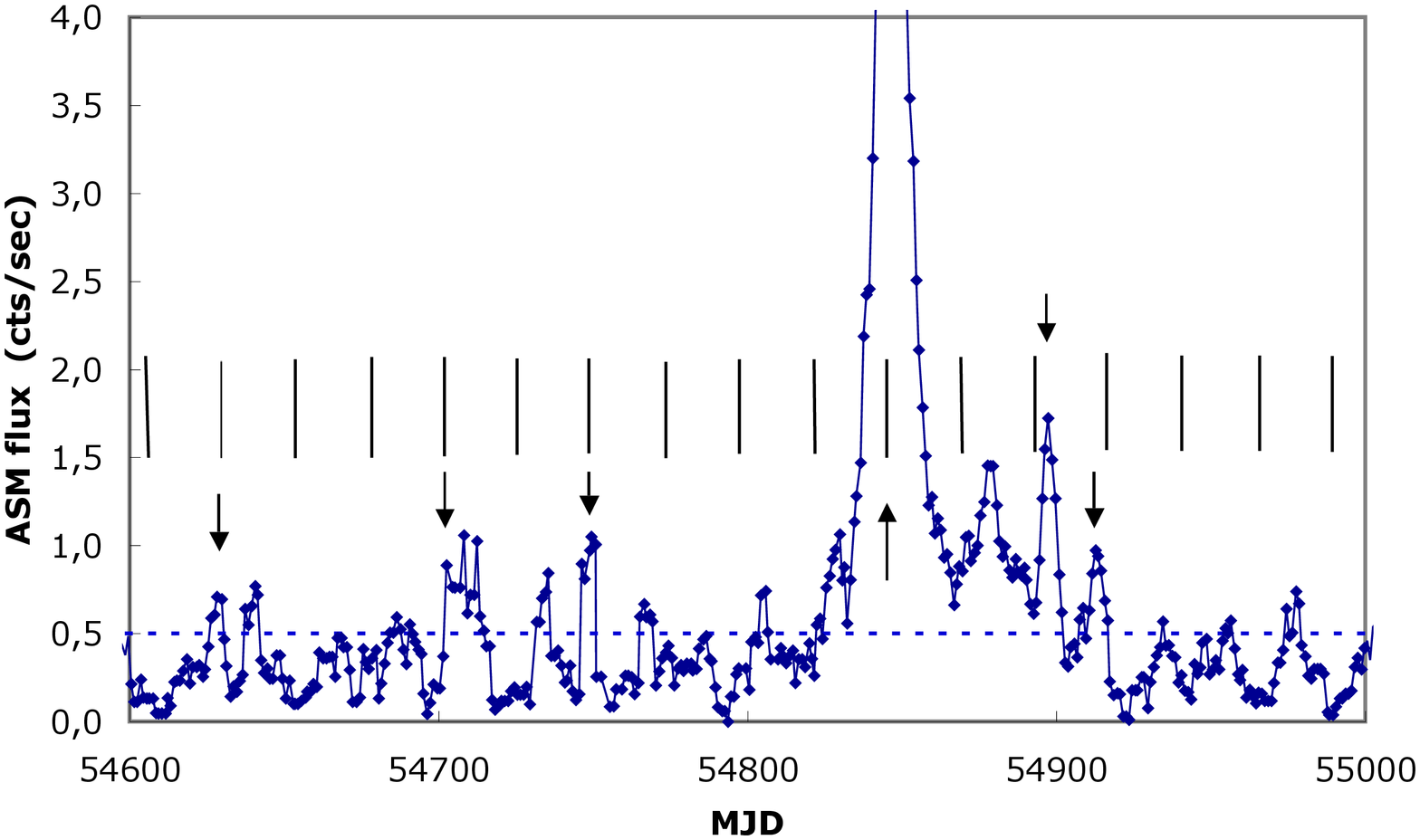}}
  \caption{A section of the X-ray light curve of 1A~1118-616
  as observed by the \textsl{RXTE}/ASM (smoothed using a running
  mean of five consecutive days). The large burst of January 2009,
  reaching a peak flux of 7.6\,cts/s at MJD 54845.4 is included. 
  The short vertical lines are at phase 0.0 with respect to the ephemeris 
  with P$_{\rm orb}$ = 24.012\,d and the arrows point to those small flares with
  peak fluxes $>$0.5\,cts/s (dotted horizontal line), which happen close 
  to phase 0.0 (several others appear to happen close to phase 0.5).}
  \label{fig:ASM_lc}
\end{figure}

As already mentioned, the source remains particularly active after the 
second as well as after the third large outburst for several tens of days. 
Fig.~1 in \citet{Coe:1994p3488}  
shows this for the second large burst (observed by \textsl{CGRO}/BATSE in 
January 1992): there is highly structured X-ray emission with several peaks,
the largest of which are around $\sim$26\,d and $\sim$49\,d after
the peak of the large burst (with peak fluxes $\sim$0.5  and  $\sim$0.4 of the 
big burst, respectively). In Fig.~\ref{fig:ASM_lc}, we show the light curve in the 
vicinity of the third large burst (of January 2009) as observed by \textsl{RXTE}/ASM
(in a smoothed version of the daily light curve with a running mean of
five consecutive days). Three peaks can also be identified within $\sim$70\,d 
of the the large burst, none of which corresponds exactly to phase zero,
but the second and the third one are close (the mean of the three
separations - starting with the large burst - is $\sim$23\,d).

\subsection{Timing of small X-ray flares}

\begin{figure}
  \resizebox{\hsize}{!}{\includegraphics[angle=90]{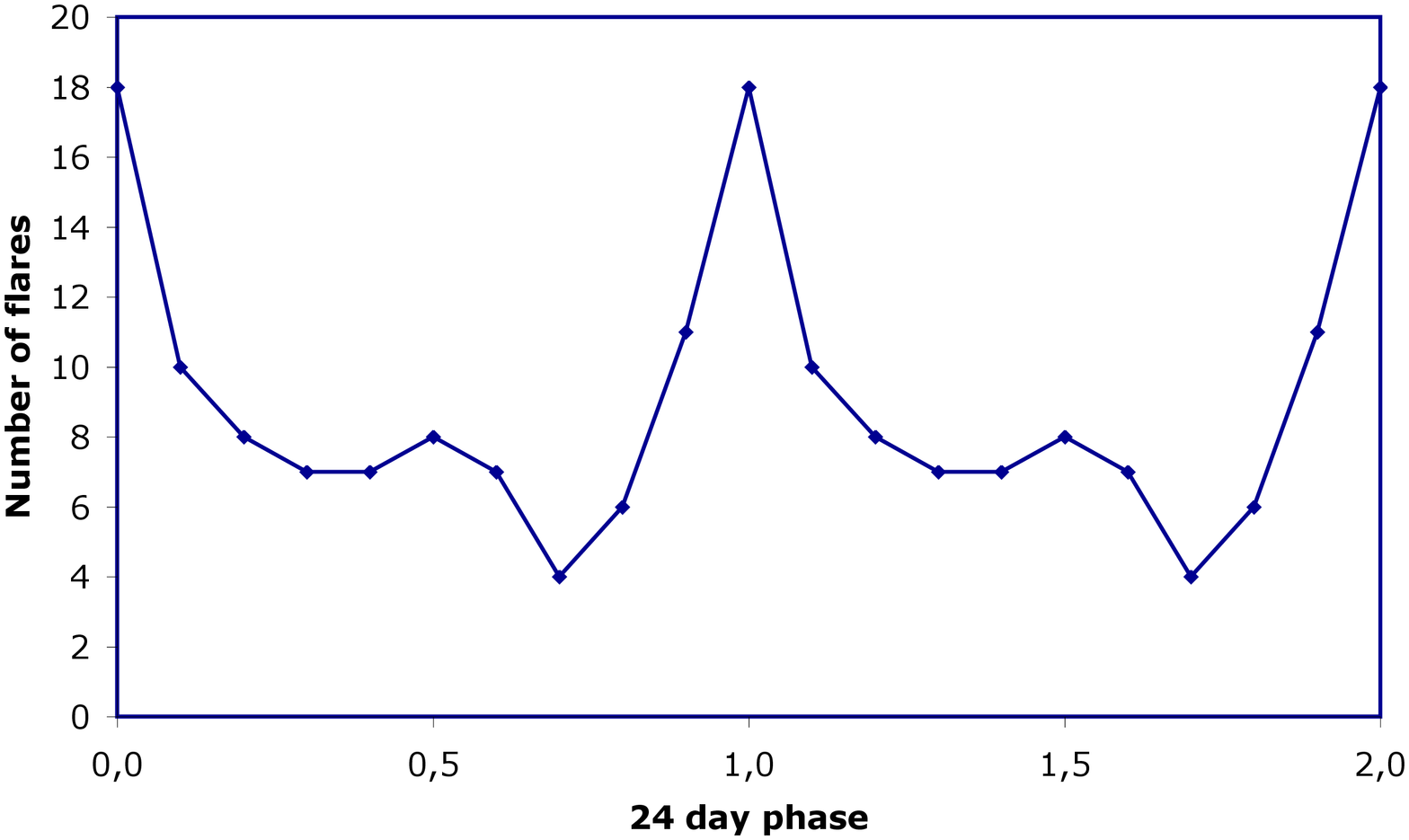}}
  \caption{Frequency histogram of ASM small flares with peak flux $>$0.5\,cts/s
  (from smoothed daily light curves) as a function of the 24.012\,d phase.}
  \label{fig:frequ}
\end{figure}

In the section of the  ASM light curve shown in Fig.~\ref{fig:ASM_lc}, 
three small peaks before the big burst (highlighted by small arrows) happen
quite close to phase zero. There are also flares at other phases, but 
counting the number of flares with peak fluxes higher than 0.5\,cts/s
over the complete ASM light curve (MJD 50087 - MJD 55315), we find 39 out 
of 86 flares (45\,$\%$) between phase  -0.15 and +0.15. The rest are almost 
uniformly distributed with a slight enhancement around phase 0.5. The analysis 
of excursions with fluxes $>$1.0\,cts/s confirms this result: 15 out of 32 events 
between phase -0.15 and +0.15.
Fig.~\ref{fig:frequ} shows the frequency distribution for the case of the 
$>$0.5\,cts/s peaks. When the same analysis is repeated with orbital
periods corresponding to separations between the large outbursts other
than 259 orbital cycles (that is between 255 and 263), the rate of coincidences 
of small flares with phase zero is significantly less. We take this as an 
indication, not proof, that 259 is probably the right number.
  
When a simple epoch folding of the long-term light curve of 1A~1118-616
as observed by \textsl{RXTE}/ASM is done, no significant peak is found
at $\sim$24\,d. This is not inconsistent with the above considerations 
of medium size and small X-ray flares: there are only very few (not well
aligned) medium size flares, occurring after the big outbursts, and the small 
flares ($>$0.5\,cts/s) have such low fluxes that both types of flares 
are completely buried in the "noise" represented by the daily flux 
measurements $<$0.5\,cts/s. When a \textsl{dynamical power 
density spectrum} (PDS) (see e.g. \citealt{Wilms_etal01}) is generated
(with an individual data set length of 2000\,d and a step size of 10\,d),
a weak signal in the form of a broad peak between 22\,d and 25\,d is found.
The PDS is capable of detecting quasi-periodic signals allowing for
frequency variations with time.

\section{Inclination and mass function}

Using the orbital elements determined above and an estimate of the
mass of the optical companion, we can constrain the inclination of
the binary orbit of 1A~1118/He 3-640.  The optical companion was 
classified by \citet{JanotPacheco:1981p3486} as an O9.5 III-Ve star 
with strong Balmer emission lines and an extended envelope.
\citet{Motch_etal88} assumed a mass of 18  $M_{\odot}$  for the 
optical companion. When the calibration of O-star parameters of 
\citet{Martins_etal05} is used, the value of 18 $M_{\odot}$ is confirmed
(interpolating the values for O9.5 of Tables 4 and 5 of \citealt{Martins_etal05}). 
Adopting this value and assuming 1.4 $M_{\odot}$ for the mass of the neutron 
star, Kepler's third law leads to a physical radius of the (circular) orbit of
a = 219.1\,light-s. The observed a~sin~i = 54.85\,light-s then
leads to sin~i = 0.25 and to the inclination of $i = 14.5^{\circ}$. This value 
is consistent with no eclipses being observed. The mass function is 
f(M) = (m$_{2}$~sin i)$^{3}$ / (m$_{1}$+m$_{2}$)$^{2}$ = $1.14 \times 10^{-4}$.

\section{Discussion}

The detection of the period of 24\,days for the binary orbit of 
1A~1118-616 rests mainly on the observation and analysis of
orbital delays of the arrival times of the X-ray pulses (with a 
pulse period of $\sim$407\,s). The main data set is from observations 
by \textsl{RXTE}/PCA, which sampled the large X-ray outburst of 
January 2009. Fortunately, the duration of the outburst as well as the
length of the observations were long enough to cover slightly more 
than one complete orbit. It was also fortunate that the sampling 
pattern was dense enough to allow pulse phase connection between 
the sampling intervals. However, with data for only one orbit the determination of the 
orbital period using these data has a rather large uncertainty of about $\pm 0.4$\,d.
We asume this value of 0.4\,d as the final uncertainty in the orbital
period,  despite the evidence from the small flares that 24.012\,d,
corresponding to a separation of 259 orbital cycles between the three 
large outbursts, may be the correct value.  

For the January 2009 outburst, we have found that the peak of the X-ray
flux coincides with T$_{\pi/2}$, i.e., an orbital longitude of $90$\,$^{\circ}$, while
the formal value of $\Omega$ was determined to be $310$\,$^{\circ}$.
This supports our reservation of taking the formally achieved 
$\epsilon$ / $\Omega$ combination seriously: an F-test does give a 
rather high probability ($\sim$22\,$\%$) that the improvement in $\chi^{2}$ 
(when these two free parameters are introduced) is just by chance.
On the other hand, the finding that a majority of the small flares and
one of the large bursts do occur at or around phase zero of
our 24.012\,d ephemeris indicates that the eccentricity may be somewhat
larger than zero, but most likely $<$0.16.

For completeness, we mention here that in optical observations 
of He~3-640/Wray~793 large values of H$_\alpha$ equivalent width are 
occasionally observed. \citet{Coe:1994p3488} had found a value 
in excess of 100\,Angstrom on MJD 48706 (80\,d after the peak of
the January 1992 large X-ray outburst) and they associated the two 
phenomena with each other, which is clearly justified by examples
of these associations in other Be X-ray binaries
\citep{Grundstrom07, Kizilo09, Reig10}.

With respect to the timing of the three observed large X-ray bursts,
we emphasize that the two separations are the same at 17.04\,yrs. 
\citet{Motch_etal88} already concluded that the envelope of 
He~3-640 is probably not in a stationary state but undergoes
expansion and contraction phases on a time scale of several years."
We suggest that the 17\,yrs between the large X-ray outbursts is
a characteristic period for the oscillation of the envelope of He~3-640. 

Finally, we look at the position of 1A~1118-616 in the Corbet-diagram,
which relates the orbital period to the spin period (see e.g., 
\citealt{Reig04,Rodriguez09}). For Be binaries, there is considerable
scatter around a mean correlation trend, which was quantified by
\citet{Corbet86} by the following formula: 
 P$_{\rm orb}$ = 10~days $\times$ $(1-e)^{-2/3}$ (P$_{\rm spin}/1~s)^{1/2}$.
According to this formula, an orbital period of $\sim$200\,d or more
had been expected (e.g., \citealt{Motch_etal88}), a prediction that
in addition to the generally low level of the X-ray flux and the rareness 
and shortness of the larger outbursts may have helped the 24\,d orbital 
period escape detection for 35\,yrs after the source's discovery in 1975.
1A~1118-616 is indeed at the edge of the Be star distribution towards
the region where wind-fed SGXRBs (super giant X-ray binaries)
accumulate, and the second most offset Be star after SAX~J2103.5+4545.
For the latter, \citet{Reig04} had discussed a physical reason for the slow
spin at a rather short orbital period, namely that long episodes of quiescence 
in Be X-ray systems can cause a spin down to an equilibrium period that
is expected for wind-driven accretion. This may also apply to 1A~1118-616
and is consistent with the observed moderate spin-up rate ($\sim$1.1\,Hz s$^{-1}$)
during its outbursts in January 2009. A discussion of various alternative ways of
reaching very long spin periods is given in \citet{Farrell_etal08} in the context
of investigating the extremely slow ($\sim$2.7\,h) pulsar 2S~0114+650.

\begin{acknowledgements}
We acknowledge the support through DLR grant 50~OR~0702. RSt likes to
thank Klaus Werner and Dima Klochkov for discussions about the optical 
companion and the analysis of the ASM light curve, respectively. 
RER and SS acknowledge the support under NASA contract NAS5-30720.
We thank Robin Corbet for pointing us to SAX~J2103.5+4545 and for his 
data collection of the Corbet-diagram.
\end{acknowledgements}

\bibliographystyle{aa}
\bibliography{refs_auto_clean}

\end{document}